\documentclass[pra,showpcs,amsmath,amssymb]{revtex4-1}

\usepackage{graphicx}
\usepackage{dcolumn}
\usepackage{bm}
\usepackage{amssymb}

\begin{document}

\title{Shot-noise-limited spin measurements in a pulsed molecular beam}

\author{E. Kirilov, W.C. Campbell, J.M. Doyle, G. Gabrielse, Y.V. Gurevich, P.W. Hess, N.R. Hutzler, B.R. O'Leary, E. Petrik, B. Spaun, A.C. Vutha, and D. DeMille}
\date{\today}

\begin{abstract}
Heavy diatomic molecules have been identified as good candidates for use in electron electric dipole moment (eEDM) searches. Suitable molecular species can be produced in pulsed beams, but with a total flux and/or temporal evolution that varies significantly from pulse to pulse. These variations can degrade the experimental sensitivity to changes in spin precession phase of an electrically polarized state, which is the observable of interest for an eEDM measurement. We present two methods for measurement of the phase that provide immunity to beam temporal variations, and make it possible to reach shot-noise-limited sensitivity. Each method employs rapid projection of the spin state onto both components of an orthonormal basis. We demonstrate both methods using the eEDM-sensitive $H ~\! ^3\Delta_1$ state of thorium monoxide (ThO), and use one of them to measure the magnetic moment of this state with increased accuracy relative to previous determinations.       
\end{abstract}
\newpage
\maketitle

\section{\label{sec:1}Introduction}

Measuring the electron electric dipole moment (eEDM), $d_e$, with sensitivity to $d_e<10^{-27} e\cdot$cm would provide an improved probe of CP-violation originating at energy scales $E$ in the range  $E>1$ TeV \cite{EDMreview}.  One promising approach to search for the eEDM is to use a heavy polar molecule \cite{Sandars67}, since here the very large effective intramolecular electric field $\mathcal{E}_{mol}$ acting on the EDM enhances the measurable signal.  However, only recently have experimental techniques been developed that can provide sufficiently large molecular signals to make such measurements competitive with earlier experiments based on atomic beams \cite{Hindsnature}.  In particular, there have been significant increases in the flux available in beams of heavy, refractory and/or free radical species of the type needed for eEDM measurements \cite{nick,barry}.  Sufficient flux is important because the best possible signal-to-noise ratio ($SNR$) is determined by shot noise:  that is, for $N$ detected molecules, the maximum $SNR$ is given by $\sqrt{N}$.  However, large signal size does not guarantee good signal-to-noise; a wide variety of technical noise sources can lead to $SNR \ll \sqrt{N}$.  The primary result of this paper is the demonstration of a nearly shot-noise-limited measurement of molecular signals, of the type relevant to an eEDM measurement, in a high-flux pulsed molecular beam. 

Most EDM experiments are based on measurement of the spin precession phase (or, equivalently, precession frequency or energy difference between spin states) in a molecule or other particle \cite{Khriplovich97}. To accomplish this, an initial state of the form $|\Psi_i\rangle=(|J,+m\rangle+|J,-m\rangle)/\sqrt{2}$ is prepared, where $J$ is the total angular momentum and $m$ is its projection along the $z$-axis. This state corresponds to a spin orientation (for $J=1/2$) or alignment (for $J\ge 1$) along the $x$-axis.  The state evolves over time into the final state $|\Psi_f\rangle=(e^{-i\phi}|J,+m\rangle+e^{i\phi}|J,-m\rangle)/\sqrt{2}$, corresponding to an orientation or alignment rotated by angle $\phi$ about the $z$-axis. The phase $\phi$ is proportional to the Zeeman-like relative energy shift of the states. In applied, parallel magnetic ($\vec{\mathcal{B}} = \mathcal{B}\hat{z}$) and electric ($\vec{\mathcal{E}} = \mathcal{E}\hat{z}$) fields,  the phase is given by  $\phi=(-d_{e}\mathcal{E}_{mol}\!-\! \mu\mathcal{B})T/\hbar$.  Here $T$ is the time between preparation and probing of the quantum state; $\mu$ is its magnetic moment; and $\vec{\mathcal{E}}_{mol} \parallel \vec{\mathcal{E}}$ is the effective electric field acting on the eEDM within the molecule \cite{LeptonMoments2010}.  Experimentally, $d_e$ can be determined by observing the change in $\phi$ when $\vec{\mathcal{E}}_{mol}$ is reversed relative to $\vec{\mathcal{B}}$ (e.g.\ by reversing the applied external field $\vec{\mathcal{E}}$).

To determine $\phi$, the final state $|\Psi_f\rangle$ is measured in a particular basis, i.e.\ projected onto a particular state such as $|\Psi_x\rangle=(|J,+m\rangle+|J,-m\rangle)/\sqrt{2}$.  For this choice of measurement, the probability that the particle is detected is given by $P_x(\phi) = \eta |\langle \Psi_f | \Psi_x \rangle |^2 = \eta \cos^2{\phi}$, where $\eta$ is the overall detection efficiency for a particle in the experiment (including the effect of projective measurement with less than unit probability of detection).  Hence, with $N_0$ particles in the experiment, the signal arising from this measurement is $S_x = N_0 \eta \cos^2{\phi} = N \cos^2{\phi}$.  If $N$ is constant over the time between successive measurements of $\phi$, changes in $\phi$ can be detected simply from changes in $S_x$.  

This approach is used in several EDM experiments, e.g.\ in Refs.\ \cite{Hindsnature,TlEDM}, where RF excitation followed by laser-induced fluorescence is used to detect atoms or molecules in a beam. The relevant time scale for variations of $N$ is given here by the time between reversals of  $\vec{\mathcal{E}}_{mol}$, typically $\sim\!1$ s.   A closely related approach is used in Ref.\ \cite{HgEDM}, where polarization rotation of a probe laser beam is used to detect atoms in a vapor cell; and in Ref.\ \cite{paul}, where quantum beats in fluorescence are used to detect molecules in a cell. In these two cases, the phase $\phi$ advances through many multiples of $2\pi$ during a single measurement period.  In these experiments the spin precession frequency can be determined within a single measurement period; hence the relevant time scale for variations in $N$ is the spin precession period (ranging from $\sim 10^{-1}-10^{-6}$ s).

A more general method for determining $\phi$ uses projections onto \textit{both} states of an orthonormal measurement basis, e.g.\ $|\Psi_x\rangle$ and $|\Psi_y\rangle=(|J,+m\rangle-|J,-m\rangle)/\sqrt{2}$.  Noting that $P_y(\phi) = \eta | \langle \Psi_f | \Psi_y \rangle|^2 = \eta \sin^2{\phi}$ and $S_y = N_0 \eta \sin^2{\phi}$, the asymmetry $\mathcal{A}=(S_x - S_y)/(S_x + S_y) = \cos{(2\phi)}$ provides a means to determine $\phi$ that is independent of $N$, and hence insensitive to its fluctuations \cite{amar}.   This method has been used e.g.\ in neutron EDM experiments (see e.g.\ \cite{neutronEDM}), where after RF excitation the two basis states are separated by a polarizing mirror and then separately detected.    

In this work, we demonstrate methods to perform shot-noise-limited measurements of $\phi$ in a pulsed molecular beam.  We use a new type of beam source that delivers unprecedented flux, but suffers from significant variations in $N$ across a wide range of time scales \cite{nick,barry}.  Our measurements are performed on the $H\, ^3\Delta_1$ state of thorium monoxide (ThO). This state has been identified as a promising system for detecting the eEDM \cite{amar, Meyer08} due to its $\Omega$-doublet energy level structure (which provides both high eEDM sensitivity and powerful means to reject systematic errors \cite{ComminsFestschrift, Kawall04}), its small magnetic moment \cite{Brown03,hmoment} (which suppresses sensitivity to magnetic noise and systematics \cite{cornell1,Meyer08}), and its long lifetime \cite{amar,cornell1,Meyer06} (which enables high sensitivity to the phase $\phi$).   We show two separate methods that allow projective measurements onto both states of the measurement basis for each particle, making it possible to form the $N$-independent asymmetry $\mathcal{A}$ in both cases.  Finally, we use one of these methods (with some additional features) to make an improved measurement of the magnetic moment $\mu_H$ of the $H$ state of ThO.

\section{\label{sec:2} Basic approach}

All measurements here are performed on the $\Omega$-doublet of $J\! =\! 1$ rotational states in the $H ^3\Delta_1$ state of ThO, in the presence of parallel $\vec{\mathcal{E}}$ and $\vec{\mathcal{B}}$ fields defining the $z$-axis.  The energy level structure of this system is essentially the same as that described in Refs.\ \cite{paul,amar}; we refer the reader to those papers for a more detailed discussion of the system's properties, and use the notation of \cite{amar}.  In short: the $\mathcal{E}$-field is sufficiently strong ($\mathcal{E}\!\approx\! 100$ V/cm) that the $|J\! =\! 1,|m|\! =\! 1\rangle$ sublevels are (to an excellent approximation) fully polarized.  These states can be written in the form $|J\! =\! 1,m,\mathcal{N}\rangle$, where $\mathcal{N}\! =\! \pm 1$ is an approximate quantum number describing the polarization of the molecule along or against $\vec{\mathcal{E}}$, respectively. The polarized states (with defined $\mathcal{N}$) are complete mixtures of the opposite-parity $\Omega$-doublet states that are energy eigenstates in the absence of $\vec{\mathcal{E}}$. In terms of parity eigenstates $|J\! =\! 1,m,P\rangle$ (where $P\! =\! \pm 1$ is the parity eigenvalue), we can write 
\begin{eqnarray}
|J\! =\! 1,m,\mathcal{N}\rangle \! &=&\! ( |J\! =\! 1,m,P\! =\! +1\rangle \nonumber  \\
&-& \mathcal{N}\cdot sgn(m)|J\! =\! 1,m,P\! =\! -1\rangle )/\sqrt{2}. \label{Nstate_in_terms_of_Pstates}
\end{eqnarray}
 In this paper we work entirely with the states where $m\! =\! \pm 1$ and $\mathcal{N}\! =\! -1$ (though both $\mathcal{N}$ states are used in the eEDM experiment).  Due to the tensor Stark shift, the $|J\! =\! 1,m\! = \! \pm 1,\mathcal{N}\! = \! -1\rangle$ states are shifted above the $|J\! =\! 1,m\! =\! 0\rangle$ sublevels by an amount large compared to the energy resolution of the experiment (see Fig.\ 1a). To simplify notation going forward, we suppress electronic and rotational state labels and write the relevant states simply as $|\mathcal{N},m\rangle$.  In analogy with the previous discussion, we define for the remainder of the paper $|\Psi_{x/y}\rangle=(|\mathcal{N},+m\rangle\pm |\mathcal{N},-m\rangle)/\sqrt{2}$ as the measurement basis states.

The $|\mathcal{N},m\rangle$ states are probed by laser excitation to $J\! =\! 1$ sublevels of the short-lived $C ~ ^1\Pi_1$ state, and detection of subsequent fluorescence from the decay $C\!\leadsto\! X$.  The basic probing scheme is shown in Fig.\ 1.  The $C$ state also has $\Omega$-doublet structure; here both the $\Omega$-doublet splitting ($\Delta_C \approx\! 2\pi\times 51$ MHz for $J\! =\! 1$ \cite{Edvinsson65}) and the tensor Stark shift are sufficiently large that the $|C; J\! =\! 1,m\! =\! 0; P\! =\! \pm 1\rangle$ sublevels are spectrally resolved by the probe laser.  (Note that these $m=0$ parity eigenstates do not mix with each other, because the relevant Clebsch-Gordan coefficient vanishes.) To simplify notation, going forward we denote these states simply as $|C;P\rangle$. The applied $\mathcal{B}$ field is sufficiently small that the Zeeman shift between the $|\mathcal{N},m\! =\! \pm 1\rangle$ states is \textit{not} resolved by the probe laser.  We probe with linearly polarized light (polarization $\hat{\epsilon}_{probe} \! =\!  \hat{x}$ or $\hat{y}$) resonant with the $|\mathcal{N}\! =\! -1,m \! =\!  \pm 1\rangle \rightarrow |C; P\! =\! +1\rangle$ or $\rightarrow |C; P\! =\! -1\rangle$ transition.  This probe arrangement leads to exactly the type of projective measurements described above.  For example, with $\hat{\epsilon}_{probe} \! =\!  \hat{x} (\hat{y})$ and excitation to $|C; P\! =\! +1\rangle$, only the superposition state $|\Psi_{y}\rangle (|\Psi_{x}\rangle)$ is detected; $|\Psi_{x}\rangle (|\Psi_{y}\rangle)$ is a dark state and is unaffected by the probe (see e.g.\ \cite{BKDbook}).  Hence, projective measurements of both basis states can be performed by probing with both polarizations.  This method is applicable to many atomic and molecular systems. 

\begin{figure}
\centering
\includegraphics[width=83mm]{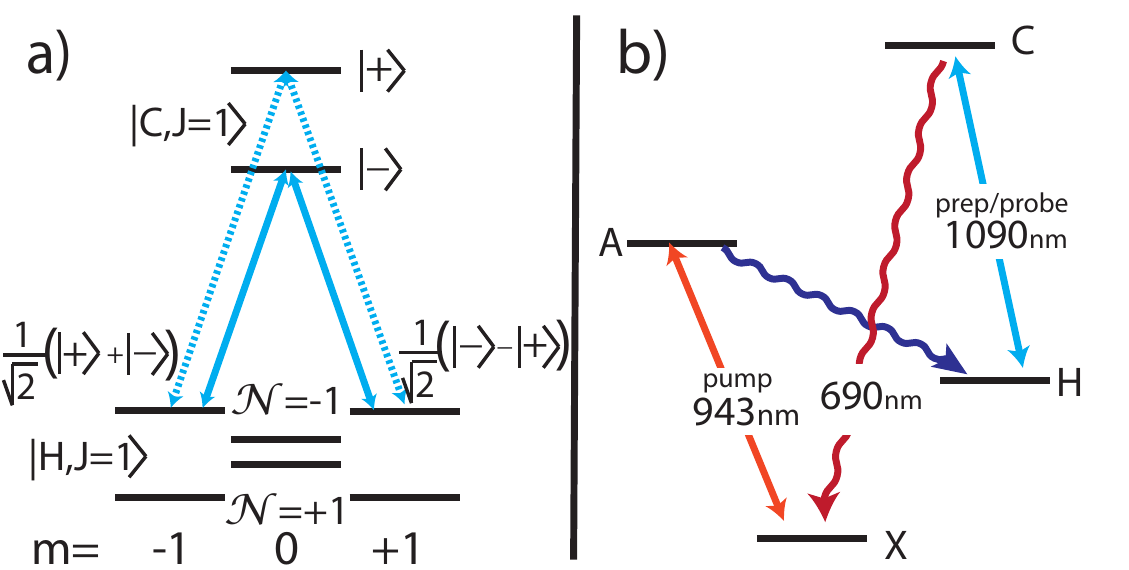}
\caption{(color online)  Relevant optical transitions.  a) Energy level structure of the $H$ and $C$ state sublevels relevant to this paper.  Arrows denote probe transitions.  The Stark-shifted $m\!=\!\pm 1$ sublevels of the $C, J\!=\! 1$ state are not shown. Here the ket $|\pm\rangle$ denotes parity. b) Electronic states and transitions used for state population, preparation, and probing.  Straight (wiggly) arrows correspond to stimulated absorption (spontaneous emission).
}\label{Fi:AOMs}
\end{figure}

Another method for projective measurements onto both spin basis states, which we refer to as ``probe-parity switching'', takes advantage of some specific behavior of molecular $\Omega$-doublets in the presence of a completely polarizing $\mathcal{E}$-field. In particular, the $|\mathcal{N}\! =\! -1,m\! =\! \pm 1\rangle$ sublevels are each a balanced superposition of pure parity states, but the relative sign of the superposition amplitudes is opposite for $m\! =\! \pm 1$ (see Eqn.\ \ref{Nstate_in_terms_of_Pstates} and Fig.\ \ref{Fi:AOMs}a).   Because of this, excitation to the $|C; P\! =\! -1\rangle$ state rather than $|C; P\! =\! +1\rangle$ interchanges the roles of $|\Psi_{x}\rangle$ and $|\Psi_{y}\rangle$ for a given probe polarization \cite{paul}.  (We note that this subtlety is not considered in Ref.\ \cite{amar}, where the ThO system was introduced.)  Hence, it is also possible to perform projective measurements of both basis states, by keeping the polarization fixed and probing on the two transitions $|\Psi_f\rangle \rightarrow |C; P\! =\! \pm 1\rangle$.  We demonstrate both the polarization and probe-parity projection methods here.

In principle, it is possible to perform projective measurements on both states for each molecule in the beam by using two different detection regions with different probe configurations in each.  Instead, we use a single detection region and subject each molecule to both probe conditions within this region.  This is accomplished by rapidly switching between polarizations or excited-state parities (the latter via changing laser frequency), with period shorter than the time it takes for molecules to traverse the probe laser beam.  Time-resolved detection is then used to distinguish the signal under one probe condition from that under the other.

Population and preparation of the initial state $|\Psi_i\rangle \! =\!  |\Psi_{x}\rangle$ is accomplished as follows (see Fig.\ 1). We excite population from the ground state $|X;J\!=\!1,m\!=\!\pm 1\rangle$ sublevels to the $|A;J\!=\!0\rangle$ state using laser light at 943 nm wavelength.  We refer to this as the pump laser. The excited state spontaneously decays to an incoherent superposition of $m$ state sublevels in the $|H,J\!=\!1\rangle$ manifold. Next, the molecules traverse a linearly polarized laser beam (with polarization $\hat{\epsilon}_{prep} \! =\!  \hat{x}$), which drives the $|\mathcal{N},m\! =\! \pm 1\rangle\rightarrow |C;P\! =\! +1\rangle$ transition at 1090 nm wavelength. This laser (referred to as the state preparation laser) pumps out the ``bright'' state $|\Psi_{y}\rangle$ and leaves behind population in the ``dark'' state $|\Psi_{x}\rangle$ used as our pure initial state: $|\Psi_{i}\rangle = |\Psi_{x}\rangle$.

Following preparation of $|\Psi_{i}\rangle$, molecules in the beam traverse an interaction region of length $L \approx 22$ cm and accumulate relative phase $\phi$, given by
\begin{equation}
\phi = \phi_{\mathcal{E}}+\phi_{\mathcal{B}} = -\frac{2}{\hbar}\int_{0}^{L}\left[ d_e \mathcal{E}_{mol} + \frac{\mu_{H}}{2} \mathcal{B} \right] \frac{d\ell}{u}, \label{phase_integral}
\end{equation}
where the electric ($\phi_{\mathcal{E}}$) and magnetic ($\phi_{\mathcal{B}}$) contributions to the phase correspond to the first and second terms in the integral, respectively.  Here $u \cong 180$ m/s is the forward velocity of molecules in the beam \cite{nick}, and $\mu_{H} \cong 0.008 \mu_B$ is the $H$-state magnetic moment (where $\mu_B$ is the Bohr magneton) \cite{hmoment}.  At the end of the interaction region, the final state is 
\begin{eqnarray}
|\Psi_f\rangle &=& \left( e^{-i\phi}|\mathcal{N},m\! =\! +1\rangle+e^{i\phi}|\mathcal{N}, m\! =\! -1\rangle \right) /\sqrt{2} \nonumber \\
&=& \cos{\phi}|\Psi_x\rangle -i \sin{\phi}|\Psi_y\rangle. \label{final_state}
\end{eqnarray}
This state is then probed as described above.  For maximum sensitivity to small changes, $\delta\phi$, in the value of $\phi$, the magnetic field strength is adjusted so that $\phi_{\mathcal{B}} \approx \pi/4$.  Then the asymmetry is $\mathcal{A} = \cos{(2\phi)} \approx -2\delta\phi$, where $\delta\phi \equiv \phi - \pi/4$.  From here forward we ignore the small contribution from the eEDM, and focus only on the quantities $\phi$ and $\delta\phi$. 

Both state preparation and detection laser beams extend $\delta L \cong 3$ mm along the direction of the molecular beam (and cover the $\approx\! 1$ cm transverse height of the collimated molecular beam).  Fluorescence at wavelength 690 nm, accompanying the decay $C\!\leadsto\! X$, is collected by an array of lenses and transported to photomultiplier tubes (PMTs) via fiber optic bundles.  The overall detection efficiency is $\eta \sim 1$\%.

\section{\label{sec:3}Variations in molecular beam output}

We use a hydrodynamically-enhanced cryogenic pulsed beam source \cite{Patterson} based on laser ablation of a solid ThO$_{2}$ target into a buffer gas \cite{nick}. Unless otherwise noted, the results here used neon at temperature $T \approx 20$ K as the buffer gas; a few results used helium at $T \approx 5$ K.  This type of molecular beam source is subject to significant variation in total yield, as well as in the velocity distribution \cite{nick,barry}, both between succesesive pulses and within a single pulse. Moreover, changing the ablation spot is necessary periodically, since the beam yield diminishes substantially after some number of ablation shots (here typically $\sim\! 10^4-10^5$) on the same spot on the target.  Change of the ablation position is observed to affect the overall dynamics of thermalization with the cold buffer gas, resulting in changes in beam properties such as the mean forward velocity, rotational and longitudinal temperatures, and time to exit the source chamber. 

For achieving shot-noise-limited sensitivity to small changes $\delta\phi$ in the phase $\phi$, the main source of difficulty is the variation in source flux over time.  There are a few relevant time scales for this variation.  The first is fluctuations in the integrated yield $N$ between successive beam pulses (here at source repetition rate $R = 10-50$ Hz).  We typically observe pulse-to-pulse variations at the level $\delta N/N \approx 10-20$\% .  This indicates that measurements of $\phi$ using a projection onto a single basis state during each beam pulse would only be shot-noise-limited for $N \ll (\delta N/N)^{-2} \approx 100$ detected counts per pulse.  Since we typically detect $N \approx 2000$ counts per pulse (and expect much higher count rates in the future \cite{amar}), this pulse-to-pulse switching approach is not adequate for use in a high-flux beam such as ours.  

\begin{figure}
\centering
\includegraphics[width=83mm]{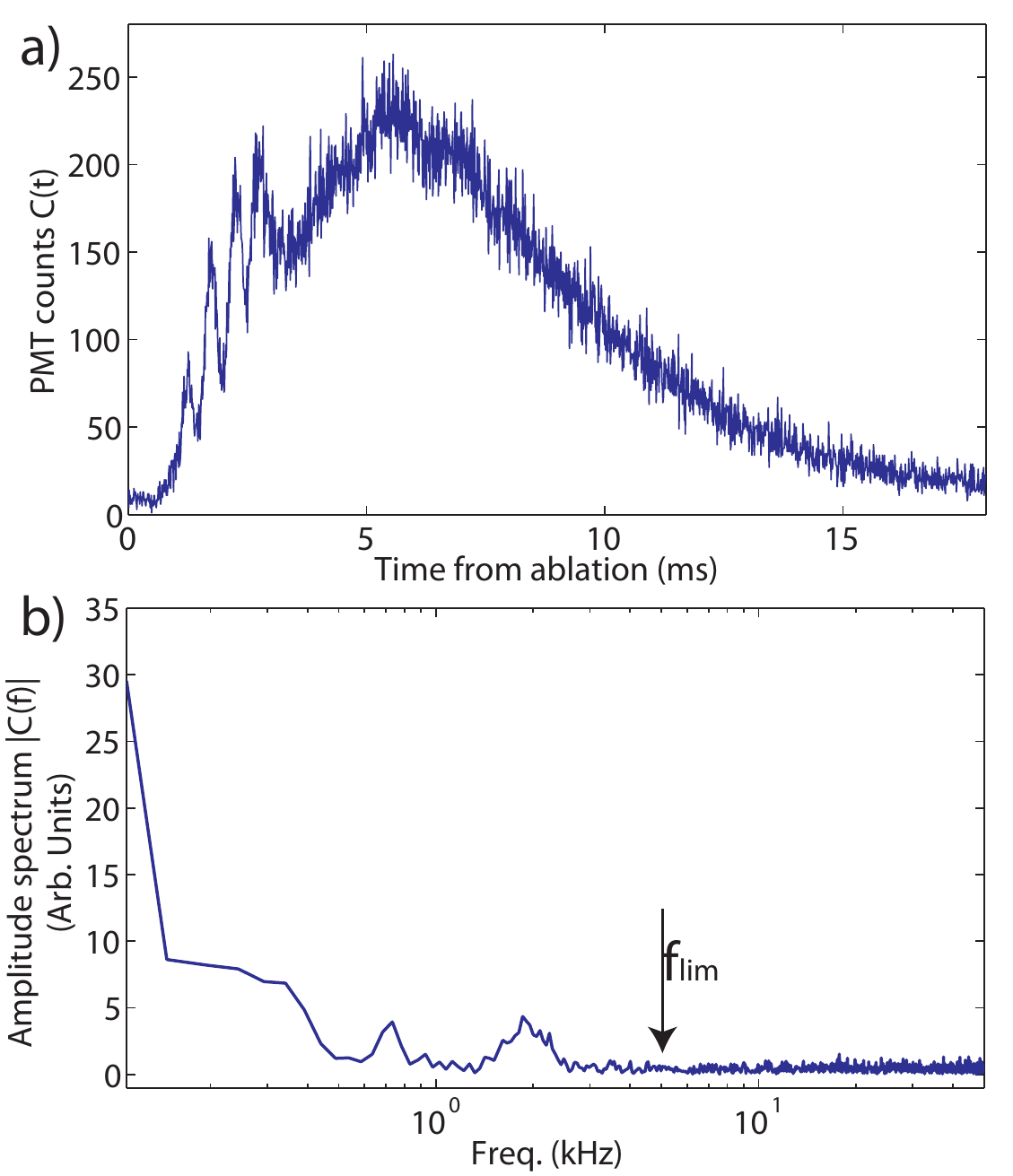}
\caption{(color online) Variations in molecular beam flux.  a) Typical temporal dependence of the ThO pulse.  Photon counts $C(t)$ in $10~\mu$s bins  (summed over $100$ ablation pulses) are plotted versus time after ablation. Here the total number of photon counts/pulse is $\sim\! 2000$. This data is taken with a helium buffer gas; with a neon buffer gas the fast fluctuations at the beginning of the pulse are typically not visible, and the pulses have shorter duration.  b) Single-sided amplitude spectrum $C(f)$ of $C(t)$. Note the flat spectrum for frequencies above $f_{lim}\! =\! 5$ kHz.}\label{Fi:pulse_shape_fourier_transform_3}
\end{figure}

Therefore we consider an approach where projective measurements onto both basis states are performed within a single beam pulse.  Again, in our system this is performed by switching between probe conditions at frequency $f_{probe}$.  Here, the minimal condition for achieving shot-noise-limited detection is that $f_{probe}$ be sufficiently large to avoid any low-frequency variations in molecular beam flux during a pulse.  Fig.\ \ref{Fi:pulse_shape_fourier_transform_3} shows the typical temporal evolution of a beam pulse, and its Fourier transform.  This shows clearly that for $f_{probe} \gtrsim f_{lim}\! =\! 5$ kHz, shot noise will be the dominant noise source.

Even better performance can be achieved by using a higher probe switching frequency.  In particular, consider a situation where each individual molecule is subject to both probe conditions.  In our experiment, this corresponds to the case where the probe period is shorter than the time required for a molecule to fly through the probe laser beam, i.e.\ $f_{probe} \! > \! u/\delta L \! = \! f_{ft} \approx 60$ kHz.  Then, since both spin projections are determined (on average) for a given molecule, the $SNR$ is entirely independent of the time scale of beam intensity fluctuations.  In addition, when $\mathcal{B}$ is adjusted to give $\phi \approx \phi_0 = \pi/4$, the probability for a molecule to be detected in a given spin projection is $\approx 1/2$.  Hence, detecting both spin projections of each molecule can increase the total signal size by a factor of 2 relative to the case $f_{lim}\! <\! f_{probe}\! <\! f_{ft}$, so that the shot-noise limit for $SNR$ improves by a factor of $\sqrt{2}$ when $f_{probe}\! >\! f_{ft}$.  This improvement requires sufficient laser power to saturate the probe transition during the time an individual probe condition holds; this is twice the power needed to saturate the signal for $f_{lim}\! <\! f_{probe}\! <\! f_{ft}$.  It also requires sufficient time resolution $\tau_d$ on the detection, such that $f_{probe} \tau_d \ll 1$.  In our case, $\tau_d$ is determined primarily by the radiative lifetime $\tau_C \approx 500$ ns of the $C$ state of ThO \cite{ourThOdata}, so this condition is easily met for $f_{probe} \gtrsim f_{ft}$.

\section{\label{sec:4}Detection by fast polarization switching}

\begin{figure}
\centering
\includegraphics[width=83mm]{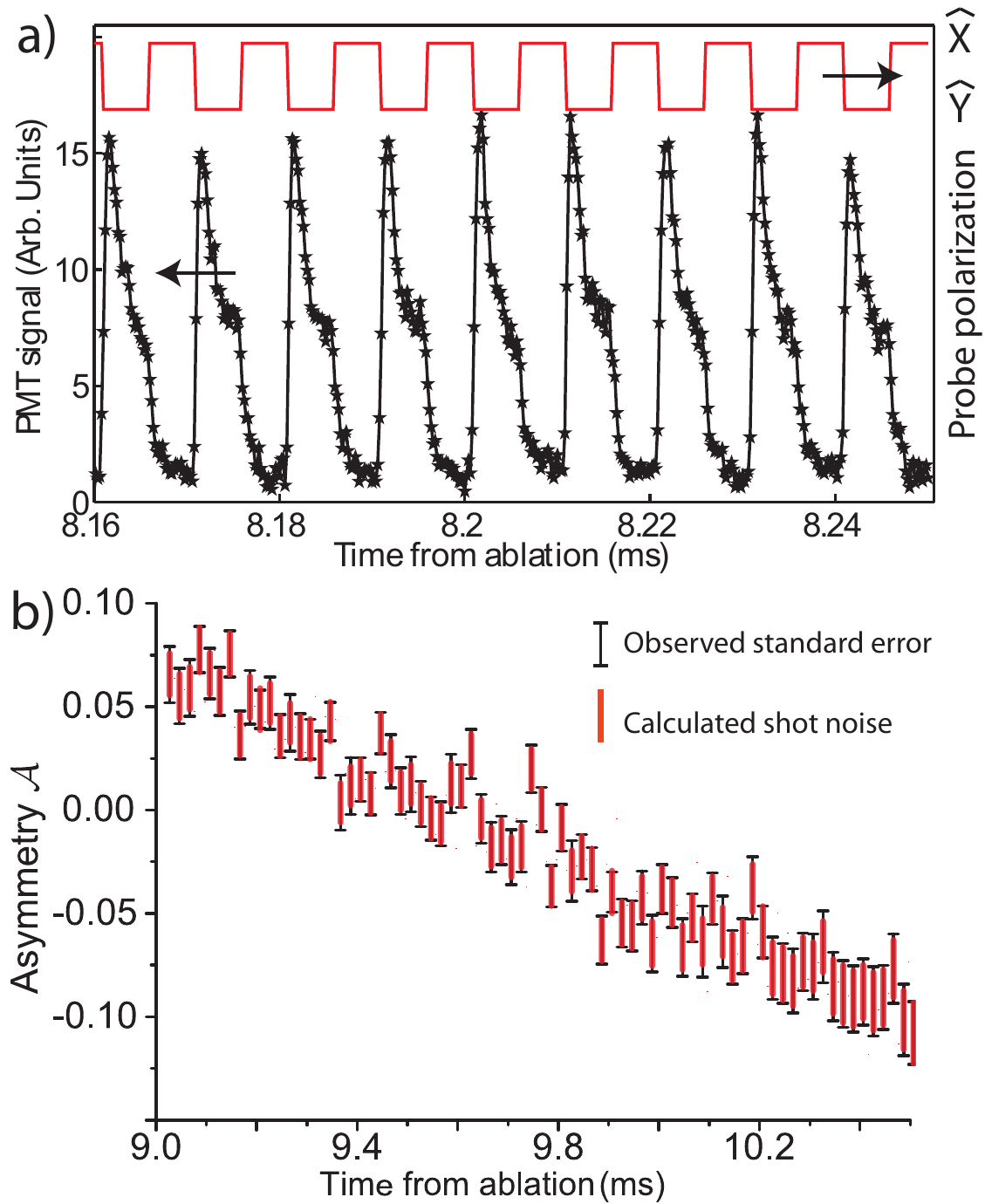}
\caption{(color online) Detection with probe polarization switching.  a) Fluorescence signal as a function of time, together with the waveform showing the time-dependent polarization of the probe laser.  Here, to make the effect of switching evident, we set $\phi \approx 0$ so that $S_x \approx 0$ and $S_y$ is near its maximum value.  b) Demonstration of shot-noise-limited detection of small phase variations. For this data, we adjust $\mathcal{B}$ to set $\phi \approx \pi/4$ as needed for optimal sensitivity to changes in $\phi$. Black bars show the measured range of uncertainty in the asymmetry $\mathcal{A}$ for each consecutive pair of polarization bins, at a given time from ablation. The evident slope in values of $\mathcal{A}$ is due to the correlation between molecular velocity and detection time (i.e., slower molecules, which undergo longer precession time $T$, arrive later). Also shown at each time is the expected uncertainty due to shot noise (red bars) calculated from the signal size in the bin pair. Bin duration is $5~\mu$s.  For clarity of presentation, only data from every other bin pair is displayed. This data is from $8\times 10^5$ molecular beam pulses, with a total number of photoelectrons $N \approx 4\times 10^8$.}\label{Fi:errorbars_vs_shotnoise_2}
\end{figure}

Our first method for detecting the spin precession phase $\phi$, proposed in Ref.\ \cite{amar}, uses switching between probe laser beams with polarization $\hat{\epsilon}_{probe} = \hat{x}$ and $\hat{y}$.  Fig.\ \ref{Fi:errorbars_vs_shotnoise_2}a shows the time-resolved fluorescence detected with polarization switching in zero $\vec{\mathcal{B}}$-field. The polarization is rapidly switched using two methods. In the first, a square-wave driven electro-optic modulator (EOM) is used to rotate the incoming polarization; in the second, two beams of orthogonal polarization are combined on a polarizing beamsplitter, with the input beams' intensities alternately modulated between on and off using acousto-optic modulators (AOMs).  We find similar performance for both methods, and do not distinguish between them in the remainder of our discussion.  The modulation frequency is $f_{probe}=100$ kHz, so $ f_{ft}\! <\!  f_{probe}\! <\! \tau_d^{-1}$ as required for optimal $SNR$. Within the period of the square wave we call each polarization state a bin.  The fluorescence signal from each bin is integrated to define the signal $S_{x/y}$ corresponding to each probe polarization. In the conditions of Fig.\ \ref{Fi:errorbars_vs_shotnoise_2}a ($\phi \approx 0$), the nearly empty bins are the ones with $\hat{\epsilon}_{probe} = \hat{x}$, i.e.\ with $\hat{\epsilon}_{probe} \parallel \hat{\epsilon}_{prep}$.

To take full advantage of the fast switching regime ($f_{probe} \! >\! f_{ft}$), sufficient laser intensity is required to ensure that each molecule within the Doppler-broadened molecular beam (with linewidth $\Gamma_D \approx 2\pi\times 2$ MHz for our collimated beam) is excited within its time of flight through the detection laser beam. Based on separate measurements of the dependence of signal size versus laser intensity on the $H-C$ transition (without polarization switching), we estimate that here (with intensity $\approx 1.5\times 10^4$ W/m$^2$), the probability for a molecule to be excited is $\gtrsim 90\%$.  The fact that we are well in the saturated regime is also apparent from the time dependence of the signal following a polarization switch (Fig.\ \ref{Fi:errorbars_vs_shotnoise_2}a). The signal in the beginning of a bin has a fast rising edge (with rise time corresponding roughly to the inverse Rabi frequency of the driven transition) and subsequently a slower decay with time constant of roughly $2 \tau_C\! \sim\! 1\, \mu$s as expected for a strongly-driven decaying system (see e.g.\ \cite{BKDbook}).  This prompt part of the signal corresponds to molecules that flew into the laser beam during the previous period of polarization, were projected onto a dark state during that previous period, and then are rapidly excited by the new polarzation as soon as it is applied. Within a bin, the signal eventually settles at an equilibrium level whose size (relative to the maximum signal in the bin) is determined by the ratio of $f_{probe}$ to the rate at which molecules emit fluorescent photons,  $(2\tau_C)^{-1}$.  We note finally that operating in the well-saturated regime also makes the signal relatively insensitive to technical noise from power fluctuations in the probe laser (though here the laser intensity noise is small enough that in any case these fluctations would contribute negligibly to our noise).

For each population residing in $|\Psi_{x/y}\rangle$ to be converted into detected photons, we must ensure that adiabatic effects are not present during the time of the polarization switch. Because the states $|\Psi_{x/y}\rangle$ and $|C; P=+1\rangle$ form a $\Lambda$-like configuration, Stimulated Raman Adiabatic Passage state transfer between the $H$-state sublevels is possible \cite{STIRAPreview98} and could result in swapping of population between $|\Psi_{x/y}\rangle$ states without a radiative decay from the $C$ state.  The two excitation pathways have time-dependent Rabi frequencies $\Omega_{x/y}(t)$ during the switching.  A general condition for nonadiabaticity is that  $d\theta(t)/dt\gg\Omega_{rms}(t)$, where $\theta(t)\! =\! \tan^{-1}\left[ \Omega_{x}(t)/\Omega_{y}(t) \right]$ and $\Omega_{rms}=\sqrt{\Omega^2_{x}+\Omega^2_{y}}$. In our setup, the linear polarization rotates continuously between states due to the action of the EOM, with the angle changing approximately linearly with time during the transition period $\tau_r$.  In this case the condition for nonadiabaticity can be written in the equivalent form $\Omega_{rms}\tau_{r}\ll 1$.  This is well-fulfilled given the values $\Omega_{rms} \sim 2\pi\times 1$ MHz (for exactly resonant molecules) and $\tau_{r} \approx 10$ ns used here.  

Next we discuss the $SNR$ for changes in $\phi$ achieved with the polarization switching method. We calculate the asymmetry $\mathcal{A} = \cos{2\phi} \approx -2\delta\phi$ for the the signal from each consecutive pair of orthogonal polarization bins (labeled by the start time $t_k$ of the $\hat{\epsilon}_{probe} = \hat{x}$ bin in the pair), for each of $n_p = 8\times 10^5$ molecular beam pulses.   For each value of $t_k$, we group the data into blocks of 50 pulses and calculate the average and standard error in the mean for the set of values of $\mathcal{A}(t_k)$ in the block.  Then the uncertainty averaged over many blocks, $\delta\mathcal{A}(t_k)$, is determined from the weighted average of this set.  

To compare the experimentally determined uncertainty to its expected level due to shot noise, a few additional factors must be taken into account.  One is the effect of background counts, which contribute to shot noise but not to the signal.  
In the current state of the experiment, backgrounds due to scattered laser light and PMT dark counts constitute $\approx\! 1/3$ of the peak fluorescence signal. In addition, there are small imperfections that lead to degradation in fringe contrast (i.e., so that the extremal values of the asymmetry are $|\mathcal{A}| < 1$).  This can arise e.g.\ due to incomplete excitation by the state preparation laser, dephasing of spins within the ensemble due to the finite spread of velocities (and hence interaction times), etc.  To account for these effects, we write the signal as $N = N_{tot} - N_b$, where $N_b (N_{tot})$ is the background (total) number of photon counts, and the experimentally observed asymmetry as $\tilde{\mathcal{A}}= C\cos{2\phi}$, where $C < 1$ is the contrast. 
Including these effects, the expected standard deviation in the experimental asymmetry, due to shot noise, is $\delta\mathcal{A}_{sn} = \sqrt{N_{tot}}/(CN)$.  In Fig.\ \ref{Fi:errorbars_vs_shotnoise_2}b we show the results of this analysis for the central $\sim\! 1.5$ ms of the beam pulse, where the signal size is largest.  Note the generally excellent agreement: we find the actual $SNR$ is only about $1.1\times$ smaller than the expected shot-noise level.

\section{\label{sec:5}Detection by probe-parity switching}

Next we demonstrate the ``probe-parity'' switching method for detecting the spin precession phase $\phi$.  In this method, both states of the measurement basis are detected by rapidly switching the probe laser frequency between resonance on the $|\mathcal{N},m\! =\! \pm 1\rangle \rightarrow |C; P\! =\! +1\rangle$ and $|\mathcal{N},m\! =\! \pm 1\rangle \rightarrow |C; P\! =\! -1\rangle$ transitions.  Details of the experimental implementation are described in Fig.\ 4a.
Here unwanted adiabatic transfer between the $|\Psi_{x/y}\rangle$ states can occur due to coupling through the other parity component of the $C$-state $\Omega$-doublet; the probability of such a transfer is $P_{a}\!\sim\! \exp{(-\Delta_C^{2}\tau_{r}\gamma_{C}/2\Omega_{rms}^2)}$ \cite{Ivanov04}, where $\gamma_C = 1/\tau_C \approx 2\pi\!\times\! 300$ kHz is the $C$ state radiative linewidth \cite{ourThOdata}, $\tau_r\! \sim\! 100$ ns is the AOM switching time, and again $\Delta_C \approx 2\pi\!\times\! 51$ MHz is the $C$ state $J=1$ $\Omega$-doublet splitting.  This probability is negligible under our conditions.  

\begin{figure}
\centering
\includegraphics[width=83mm]{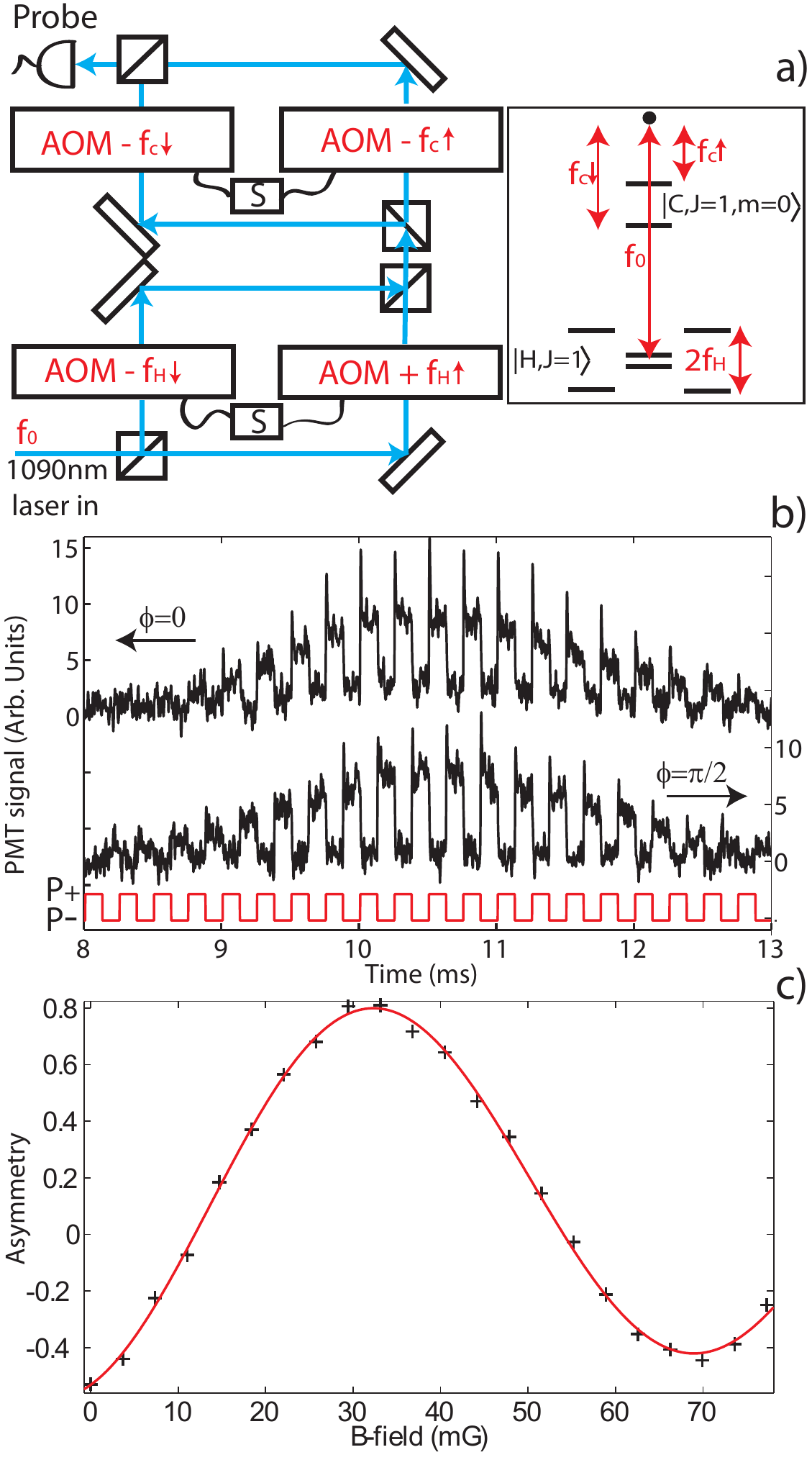}
\caption{(color online) Detection via probe-parity switching. a) Acousto-optic modulator (AOM) configuration enabling fast frequency switching between the $|\mathcal{N}, |m|\! = \! 1\rangle \rightarrow |C;P\! =\! +1\rangle$ and $ \rightarrow |C;P\! =\! -1\rangle$ transitions. (For the purposes of the eEDM measurement, this setup also makes it possible to individually address the $|\mathcal{N}\! =\!+ 1, |m|=1\rangle$ or $|\mathcal{N}\! =\!- 1, |m|=1\rangle$ sublevels of the $H$ state.)  One incoming laser, tuned to a frequency $f_{0}$ above resonance with the $C$ state, is used. In the figure, the frequency shift induced by an AOM is denoted by a +(-) sign for a positive (negative) frequency shift, and $Y\uparrow (Y\downarrow)$ indicates that the upper (lower) $\Omega$-doublet component of electronic state $Y$ is addressed. The difference between the AOM frequencies $f_{C\uparrow}$ and $f_{C\downarrow}$ is held fixed at the splitting of the $|C;P\! =\! \pm 1\rangle$ states, $\Delta_C$. The label in each AOM box indicates the frequency shift induced by that AOM, when activated.  S denotes a power switch used to set which AOM is activated. Inset: level diagram showing the combinations of shifted frequencies used to address the desired transitions. The output laser beam consists of a single frequency, shifted from $f_0$ by one of 4 possible values. This beam is sent to a saturated fiber laser amplifier and then delivered to the molecules. b) Time-resolved fluorescence from a single molecular pulse. Molecules prepared in $|\Psi_x\rangle$ are shown after evolving in magnetic fields corresponding to $\phi = 0$ and $\phi = \pi/2$. The final state is then probed by rapid switching of the probe laser frequency (with the resonant excited state parity depicted by the square wave labeled $P+$ and $P-$). The sharp increase of population in the beginning of each bin arises from the same dynamics described in the polarization switching case (Fig.\ \ref{Fi:errorbars_vs_shotnoise_2}a). c) Asymmetry $\mathcal{A}$ vs $\mathcal{B}$.  The data (points) is overlaid with the fit function (red) described in the text.} \label{Fi:C_switching_combo}
\end{figure}

Fig.\  \ref{Fi:C_switching_combo} shows data taken with this method.  In Fig.\ \ref{Fi:C_switching_combo}b, time-resolved fluorescence signals are shown for two different values of $\mathcal{B}$ and hence $\phi$, demonstrating the basic concept of spin detection via probe-parity switching. Note that for this data, the rate of switching ($f\! =\! 4$ kHz) was not sufficiently large for achieving optimal sensitivity to $\phi$; rather, this data is meant simply to demonstrate the principle of this unique method for performing orthogonal projective spin measurements in molecules with $\Omega$-doublet structure. 

We note that the probe-parity switching method differs slightly from the polarization switching method in that here the maximum signal sizes for the two different probe conditions will likely be unequal.  The origin of this difference is that the spatial pattern of fluorescence from the $|C; P\! =\! \pm 1\rangle$ states is different; hence if fluorescence is not collected from the full $4 \pi$ solid angle, the maximum signal size obtained from exciting the states is not identical.  In more detail: due to electric dipole selection rules, the $|C; P\! =\! + 1\rangle$ state decays only to $|X,J\! =\! 1,P\! =\! -1,m\!=\!\pm 1\rangle$, while the $|C; P\! =\! -1\rangle$ state decays to $|X,J\! =\! 0,P\! =\! +1,m\!=\! 0 \rangle$ and $|X,J\! =\! 2,P\! =\! +1,m\!=\!0,\pm 1\rangle$.  The branching fractions for decay to each sublevel are determined by angular factors (Clebsch-Gordan and H\"{o}nl-London).  The spatial distribution of fluorescence for $\Delta m\! =\! 0$ transitions is $p_{\Delta m \! =\!  0}(\theta,\phi) = (3/4\pi)\sin^2{\theta}$; for $\Delta m\! =\! \pm 1$ it is $p_{\Delta m \! =\!  \pm 1}(\theta, \phi) = (3/8\pi)(1+\cos^2{\theta})$ \cite{landau}.  The total spatial distribution of fluorescence from a given upper state is then a summation over the product of the these functions with the respective branching fractions. From this, we derive that the $|C; P\! =\! +1\rangle$ decays have the distribution $p_{1^+}(\theta, \phi)\propto(1+\cos^2{\theta})$, while for $|C; P\! =\! -1\rangle$, $p_{1^-} \propto (1+[23/2]\sin^2{\theta})$. The maximum signal size from each state then is proportional to the integral of these distribution functions over the angles from which fluorescence is collected.  Using a Monte Carlo simulation based on the geometry of collection optics in these experiments, we calculate the ratio $R$ of collection efficiencies ($E_{1^\pm}$ for the respective decays) to be $R = E_{1^+}/E_{1^-} \approx 1.1$.  

The net effect is that the asymmetry $\mathcal{A} = (S_x-S_y)/(S_x+S_y)$ will not be a simple sinusoid as for the case of polarization switching.  Instead, here $\mathcal{A}(\phi)=[R\cos^2{\phi}-\sin^2{\phi}]/[1 + (R-1) \cos^2{\phi}]$.  For this functional form, starting at $\mathcal{A}=1$ ($\phi=0$) at $\mathcal{B}=0$, the first zero crossings are shifted further from $\phi=0$, relative to their positions for the ideal $\cos{2\phi}$ function. For the purposes of taking EDM data this effect has little impact if, as usual, the $\mathcal{B}$-field is adjusted so that the asymmetry $\mathcal{A}\approx 0$ (the first fringe zero-crossing point) at the operating point.  Since here $\delta \mathcal{A} \propto \delta \phi$, with slope near unity, the experiment remains sensitive to small phase changes $\delta \phi$.

Fig.\  \ref{Fi:C_switching_combo}c shows a spin-rotation fringe, i.e.\ a plot of the asymmetry $\mathcal{A}$ vs. $\mathcal{B}$. As discussed earlier, the fringe contrast $C$ is expected to be smaller than unity even at $\mathcal{B}\! =\! 0$, and then even smaller as $\mathcal{B}$ increases, due to contributions from molecules with different velocities and hence different spin precession times $T$. We thus fit the data to the function 
\begin{equation}
\mathcal{A}(\mathcal{B}) = C\exp{(-p\mathcal{B}^2)} \frac{R\cos^2{(q\mathcal{B}-\phi_0)}-\sin^2{(q\mathcal{B}-\phi_0)}} {R\cos^2{(q\mathcal{B}\!-\!\phi_0)}+\sin^2{(q\mathcal{B}-\phi_0)}},
\end{equation}
with $C$, $p$, $q$, $\phi_{0}$, and $R$ as free parameters.  
The specific functional form $\exp{(-p\mathcal{B}^2)}$ used to describe the $\mathcal{B}$-dependent dephasing is appropriate when the velocity distribution is a Gaussian with width $\sigma_u$ small compared to its central value $\bar{u}$, which is a reasonable approximation here \cite{nick}. The free parameter $R$ accounts for the different photon collection efficiency of the two probe conditions; from the fit we find $R = 1.14(13)$, in good agreement with expectations. The spin-rotation fringe data matches well to this simple fit, showing that the probe-parity switching method makes it possible to detect phases of the type needed for the eEDM experiment.

We note in passing that the ``probe-parity'' method has some advantages over the polarization switching method.  The need for only a single probe polarization can enable many simplifications and improvements of the experimental design. For example, imperfections in the polarization quality (which the technology for fast polarization switching inevitably generates) are irrelevant here.  In addition, unlike for the polarization switching method, here it is not necessary for the probe laser beam to propagate along the electric field (and hence to penetrate the $\mathcal{E}$-field plates). Overall, we believe this method might prove useful for other EDM measurements that employ molecular $\Omega$-doublet states \cite{Cornell,Leanhardt}, or for future generations of the ThO experiment \cite{amar}.

\section{\label{sec:6}Application: measurement of the $H$-state magnetic moment}

In this section we describe the application of these methods to an improved measurement of the $H$-state magnetic moment $\mu_H$ (which is related to the $|H; J\! =\! 1\rangle$ magnetic moment $\mu $ via $\mu = \mu_H/[J(J+1)]$).  In principle, simple spin-precession fringe data of the type shown in Fig.\ \ref{Fi:C_switching_combo} could be sufficient to determine $\mu_H$ from the relation $\phi=\mu\mathcal{B}T/\hbar$.  However, variations of beam velocity within and between pulses made it difficult to determine the value of the interaction time $T$ with sufficient accuracy.  Hence, for this measurement we modified the experiment in a manner which made it possible to determine $T$ accurately while also achieving excellent $SNR$ on the determination of $\phi$.  The basic idea is to strobe the intensity of the pump laser that populates the $H$-state, such that a) the length in time of a single strobe is short enough so molecules excited within it can be described by a time-independent velocity distribution, and b) the time gap between consecutive strobes is long enough that molecules populated within them do not overlap in the probe region.
  
\begin{figure}
\centering
\includegraphics[width=83mm]{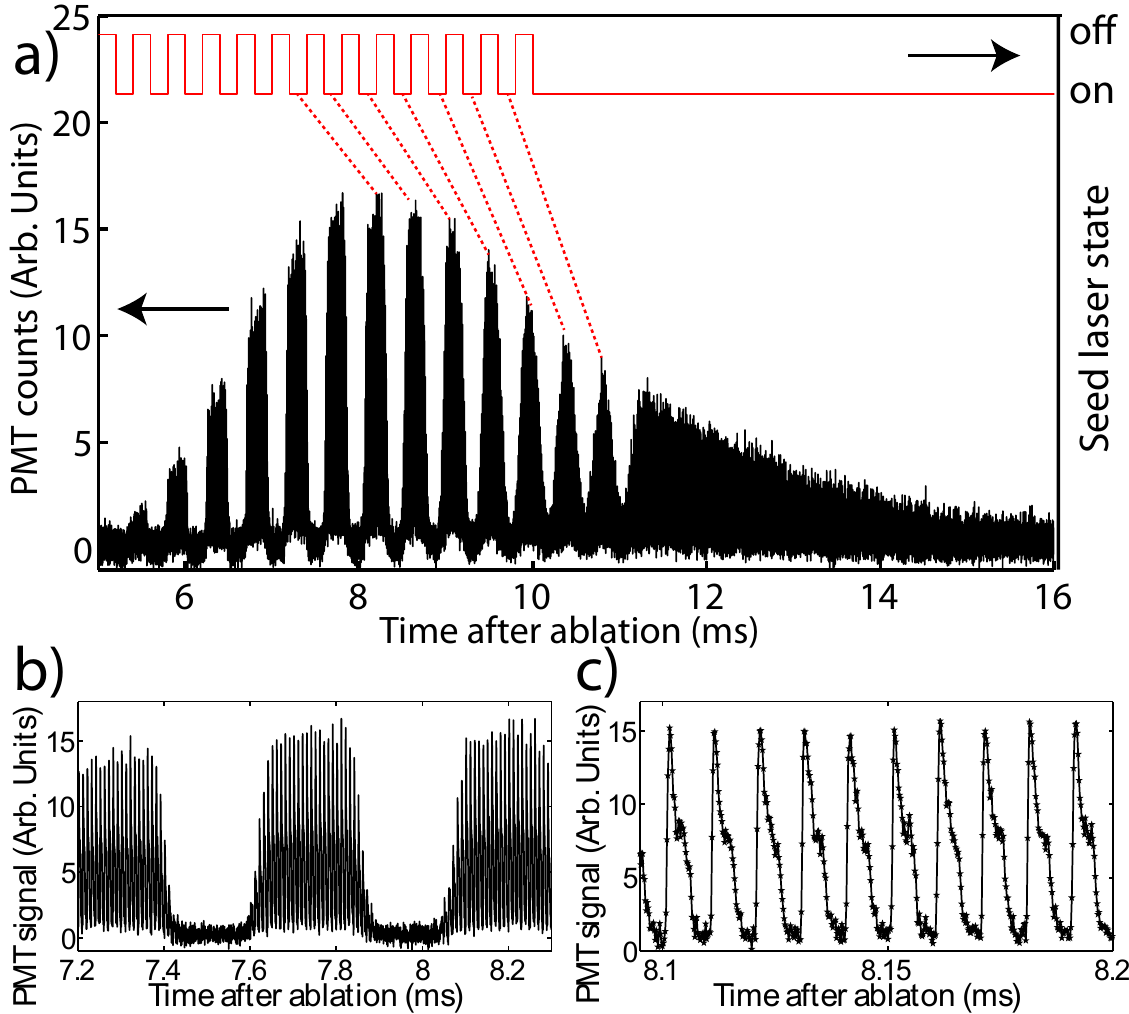}
\caption{(color online) Strobed population method for the magnetic moment measurement. a) Fluorescence signal as a function of time (averaged over 2000 molecular beam pulses), with amplitude modulation of the pump laser for temporal gating together with probe polarization switching for phase detection (see text for details). The square wave shows the pump laser amplitude time dependence. b) Zoom in to the signal from a single pump-laser strobe pulse. c) Zoom in to show the time-dependence due to probe polarization switching. }\label{Fi:3zooms_944chop_eomprobe}
\end{figure}

To implement this technique, we amplitude-modulate the pump laser that populates the $H$ state 
(see Fig.\ \ref{Fi:AOMs}b), 
as a square wave with frequency $f_{str}=2.5$ kHz and $50\%$ duty cycle.  This is accomplished with an AOM 
(Fig.\ \ref{Fi:3zooms_944chop_eomprobe}b), 
which generates a fixed number of square pulses that are phase locked to the ablation laser clock.  
Fig.\ \ref{Fi:3zooms_944chop_eomprobe}a 
shows the time-dependence of the fluorescence signal at the probe region. Here the slow modulation of the signal reveals the time structure due to strobing of the pump laser. The separate strobes clearly do not overlap in the detection region.  Within a single beam pulse, there is a monotonic increase in travel time $\tau$; the difference $\Delta \tau$ between travel times of the first and last strobe, $\Delta\tau \approx 0.1$ ms, implies a change in the average velocity of $\Delta \bar{u} \approx 26$ m/s from begnning to end of the molecular beam pulse. The velocity spread within each strobe (extracted from the fit parameter $D$ described below) is found to be much smaller, $\sigma_u \approx 5$ m/s. The combined velocity spread obtained in this manner is consistent with previous measurements on our beam source using different methods \cite{nick}.

\begin{figure}
\centering
\includegraphics[width=83mm]{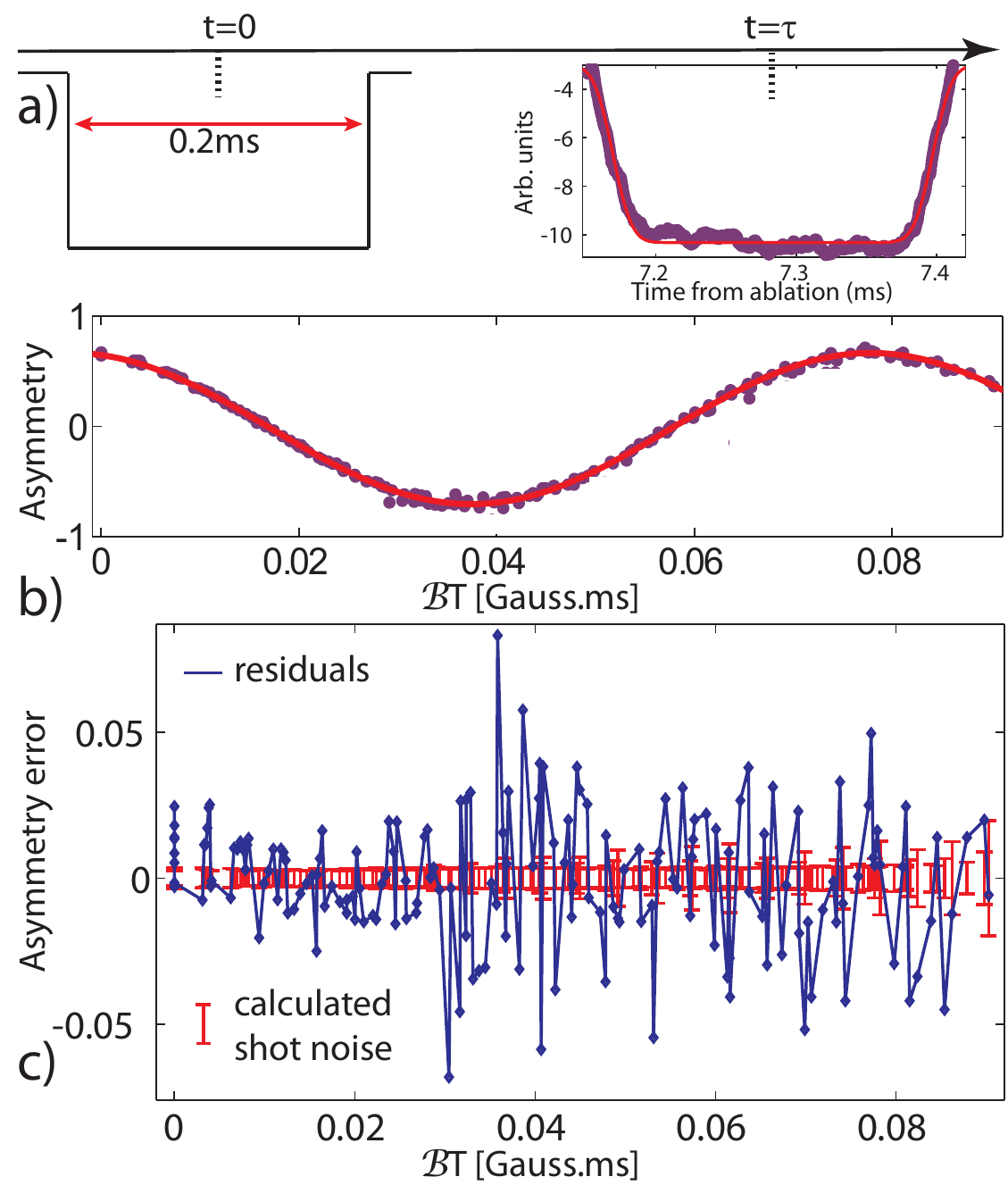}
\caption{(color online) Determination of the magnetic moment.  a) Strobe pulse (left) and subsequent detected fluorescence signal (right), along with the fit to signal as described in the text. The dashed lines show the correspondence between a given strobe pulse and the subsequent downstream fluorescence signal.  b) Plot of asymmetry vs. the product $\mathcal{B}T$ of magnetic field and precession time.  One data point is plotted for each strobe, and the curve shows a fit to a sinusoidal function. c) Fit residuals and expected level of shot noise for each strobe (after averaging over 2000 ablation shots per point).  Note that here the fit residuals are typically several times larger than the shot-noise limit; this deviation is presumably due to inadequacy of the simplifying assumptions made in the modeling of the velocity and temporal distributions of the molecules. However, the values of the residuals appear randomly distributed, so we assign no additional systematic error due to the imperfect fit.}\label{Fi:Ramsey_fit_upperH_strobe944_residuals_full_snoise}
\end{figure}

We define the position of the pump laser (a small distance $\Delta L$ upstream from the state preparation laser) as $\ell=-\Delta L$, and write the population $w_0(t)$ of $H$-state molecules passing by the position $\ell =-\Delta L$ at time $t$ as a square pulse extending from $t_{1}$ to $t_{2}$: $w_0(t)=\Theta(t-t_{1})\Theta(t_{2}-t)$, where $\Theta(t)$ is the Heaviside function.  We model the velocity distribution of the molecules within an individual strobe pulse as a Gaussian with mean velocity $\bar{u}$ and width $\sigma_{u}$.  Then, the population of $H$-state molecules arriving downstream at the position $\ell = L$ of the probe laser, at time $t$, is given by 
\begin{eqnarray}
w_{L}(t) &=& \int_{-\infty}^{\infty}\Theta(t'-t_{1})\Theta(t_{2}-t') \\ \nonumber
&\times & \frac{dt'}{(t-t')^2}\exp{ \left\{-\frac{[\frac{L+\Delta L}{\bar{u}}-(t-t')]^2} { 2[(t-t')\sigma_{u}/\bar{u}]^2 } \right\} }'. \label{arrival_dist}
\end{eqnarray}
To evaluate this expression, we keep terms of lowest order in both small quantities $\sigma_u/\bar{u}$ and $t_{str}/\tau$, where $t_{str} = t_2-t_1 = 0.2$ ms is the strobe pulse duration and the time of flight is $\tau \approx (L+\Delta L)/\bar{u} \approx 1$ ms.
With this simplification, we arrive at the analytical form $w_L(t)=A\cdot (\mathrm{erf}[(B-t)/D]-\mathrm{erf}[(C-t)/D])$ (where $\mathrm{erf}(x)$ is the error function). This function, with four free parameters ($A$,$B$,$C$ and $D$), is used for a numerical fit of the signal for each pulse arriving in the detection region (Fig.\ \ref{Fi:Ramsey_fit_upperH_strobe944_residuals_full_snoise}a). From this, the mean time of travel for molecules in each strobe pulse, $\tau=(B+C-t_1-t_2)/2$, is extracted. The quantity $\tau$ is the time between arrival in the pump and probe beams, while the free precession time $T$ of the spins is the time of flight between state preparation and probe beams.  We write $T=\xi \tau$, where $\xi = L/(L+\Delta L) = 0.94$, and relate the asymmetry to $T$ in order to extract the value of $\mu_H$.

Within each strobe, we detect the asymmetry $\mathcal{A}$ by probe-polarization switching at $f_{pr} \approx 100$ kHz  
(see Fig.\ \ref{Fi:3zooms_944chop_eomprobe}c).  
For each strobe the mean precession time $T$ is obtained from the fit and correction described above.  The procedure is repeated for values of $\mathcal{B}$ scanned in the range $0-87$ mG in 22 steps. Fig.\ \ref{Fi:Ramsey_fit_upperH_strobe944_residuals_full_snoise}b 
shows the asymmetry as a function of the product $\mathcal{B}T$ of magnetic field and precession time. This spin-precession fringe is fit to a function of the form $\mathcal{A}(\mathcal{B}T)=C\cos(\omega_{r}\mathcal{B}T+\phi_0)$, with $C$, $\omega_r$, and $\phi_0$ as free parameters. A value $\omega_{r}=77.69 \pm 0.35$ rad/(G$\cdot$ms) is derived for the precession frequency (where the uncertainty range is given as the statistical $95\%$ confidence level, with uncertainties on individual points assigned at the level of the r.m.s. fit residual).
 
 To extract the absolute $g$-factor of the molecular state, the magnetic field must be accurately calibrated. This is done by measuring $\mathcal{B}$ at many points along the axis $(x)$ of the molecular beam with a fluxgate magnetometer specified by the manufacturer to have $0.5\%$ accuracy. The uniformity of $\mathcal{B}(x)$  is found to be $\delta\mathcal{B}/\mathcal{B} \approx 10^{-3}$, so the error in the average field $\int_{0}^{L}\mathcal{B}(x)dl/L$ is dominated by the error of the magnetometer itself.  The error $\delta T$ in $T$ arises from two sources.  The first, $\delta T_v$, is due to deviations from the assumed Gaussian velocity distribution and/or square pulse intensity profile within an individual strobe, which could lead to systematic deviations from the fit shown in Fig.\ 6b and hence systematic errors in the extracted value of $T$ for any individual strobe.  This is estimated as being no larger than the product of two quantities: the total observed change in time of flight across an entire molecular beam pulse, $\xi\delta\tau$; and the fraction of the pulse covered by a single strobe, $t_{str}/t_{tot}$, where $t_{tot} = 24 t_{str}$ is the total time of observation within an individual beam pulse.  This yields $\delta T_v/T \approx 0.4\%$. An additional contribution, $\delta T_\xi$, comes from the geometric factor $\xi$ used to convert from $\tau$ (fitted) to $T$ (actual molecular precession time); from the uncertainties ($\sim 1$ mm) in the distances $L$ and $\Delta L$ we estimate these to contribute $\delta T_{\xi}/T \approx 0.4\%$.  Adding all errors in quadrature gives a final value $g_{H}=0.0088(1)\mu_{B}$. This is in agreement with, but more accurate than, our earlier determination of this value using an entirely different method \cite{hmoment}.

\section{\label{sec:7}Conclusions.}
We have presented two techniques which enable us to extract the spin precession angle of molecules in a pulsed beam at the shot-noise limit, even in the presence of much larger amplitude noise of the beam intensity. We used this method, together with a beam-chopping technique for determining molecular beam velocity, to make a measurement of the magnetic moment of the $H$ state of ThO at the $\approx\! 1\%$ level. These methods may prove useful for EDM measurements in general, and specifically for several planned eEDM measurements using molecular $\Omega$-doublet levels \cite{amar,Cornell,Leanhardt}.        

\section{\label{sec:8}Acknowledgments}

This work was supported by the NSF.
\bibliography{Ramsey_fringes_v5}
\end{document}